\documentclass[a4paper,12pt]{article}
\usepackage{epsfig,amsmath,amssymb,amsthm}

\usepackage[round]{natbib}
\usepackage[ansinew]{inputenc}
\usepackage{graphicx}

\usepackage{enumerate}
\usepackage{lscape}
\usepackage{fullpage}
\usepackage{setspace}
\usepackage{enumitem}
\usepackage{subfigure}

\usepackage{rotating}

\relax

\newcommand{\Ann}{A}
\newcommand{\ann}{}
\newcommand{\AnnNumm}{(\Ann \arabic{*}\ann)}

\newcommand{\bX}{{\bf X}}

\newcommand{\bx}{{\bf x}}

\usepackage{verbatim,color}

\def\boxit#1{\vbox{\hrule\hbox{\vrule\kern6pt
          \vbox{\kern6pt#1\kern6pt}\kern6pt\vrule}\hrule}}

\newcommand{\captionfonts}{\small}

\makeatletter
\long\def\@makecaption#1#2{%
  \vskip\abovecaptionskip
  \sbox\@tempboxa{{\captionfonts #1: #2}}%
  \ifdim \wd\@tempboxa >\hsize
    {\captionfonts #1: #2\par}
  \else
    \hbox to\hsize{\hfil\box\@tempboxa\hfil}%
  \fi
  \vskip\belowcaptionskip}
\makeatother

\begin{document}

\doublespacing

\makeatletter
\def\@seccntformat#1{\csname the#1\endcsname. }
\def\section{\@startsection {section}{1}{\z@}{-3.5ex plus -1ex minus
    -.2ex}{1.3ex plus .2ex}{\center\large\sc}}
\def\subsection{\@startsection{subsection}{2}{\z@}{3.25ex plus 1ex minus .2ex}{-1em}{\normalsize\bf}}

\def\subsubsection{\@startsection{subsubsection}{3}{\z@}{3.25ex plus 1ex minus .2ex}{-1em}{\normalsize\it}}

\newtheorem{theorem}{Theorem}
\newtheorem{definition}{Definition}
\newtheorem{lemma}{Lemma}
\newtheorem{assumption}{Assumption}
\theoremstyle{definition}
\newtheorem{example}{Example}
\newtheorem{remark}{Remark}

\parindent 0cm

\bibliographystyle{ecta}

\begin{center}
\Large{A fluctuation test for constant Spearman's rho \\ with nuisance-free limit distribution} \\
\normalsize

\vspace{0.5cm}
by
\vspace{0.5cm}

\renewcommand\thefootnote{*}

\renewcommand{\baselinestretch}{1.0}
\small\normalsize {\bf Dominik Wied \footnote{Corresponding
author. Phone: +49/231/755
5419, Fax: +49/231/755 5284.}}\\

\vspace{0.2cm}

Fakult\"at Statistik, TU Dortmund\\
44221 Dortmund, Germany\\
wied@statistik.tu-dortmund.de\\

\vspace{0.4cm}

{\bf Herold Dehling}\\

\vspace{0.2cm}

Fakult\"at f\"ur Mathematik, Ruhr-Universit\"at Bochum \\
44780 Bochum, Germany\\
herold.dehling@rub.de \\

\vspace{0.4cm}

{\bf Maarten van Kampen}\\

\vspace{0.2cm}

Ruhr Graduate School in Economics and Fakultät Statistik, TU Dortmund \\
44221 Dortmund, Germany \\
maarten.vankampen@tu-dortmund.de \\

\vspace{0.4cm}

and

\vspace{0.4cm}

{\bf Daniel Vogel}\\

\vspace{0.2cm}

Fakult\"at f\"ur Mathematik, Ruhr-Universit\"at Bochum \\
44780 Bochum, Germany\\
vogeldts@rub.de \\
\vspace{0.4cm}
This version: \today
\end{center}

\newpage

\begin{abstract}
A CUSUM type test for constant correlation that goes beyond a
previously suggested correlation constancy test by considering Spearman's rho in
arbitrary dimensions is proposed. Since the new test does not require the existence
of any moments, the applicability on usually heavy-tailed financial data is greatly improved.
The asymptotic null distribution is calculated using an invariance principle for the
sequential empirical copula process. The limit distribution is free of
nuisance parameters and critical values can be obtained without bootstrap techniques.
A local power result and an analysis of the behavior of the test in small samples is provided.

\end{abstract}


\newpage

\section{Introduction}
Recently, Wied, Krämer and Dehling (2012)\nocite{wied:2012} proposed a fluctuation test for constant correlation based on the Bravais-Pearson correlation coefficient.
The test, which will be referred to as BPC test in the following, is for example useful in financial econometrics to examine changes in the correlation between asset returns over time.
\citet{longin:1995} and \citet{krishan:2009} discuss the relevance of this question. The test complements former approaches by e.g.\ Galeano and Pe\~na (2007)\nocite{galeano:2007} and \citet{aue:2009}.
However, one major drawback of this test is the fact that the limit distribution is derived under the condition of finite fourth moments (similar to \citealp{aue:2009}).
This is a critical assumption
because the existence of fourth moments in usually heavy-tailed financial returns is doubtful, see e.g. \citet{grabchak:2010}, \citet{kraemer:2002} and \citet{amaral:2000}.

This paper presents a fluctuation test for constant correlation based on Spearman's rho which imposes no conditions on the existence of moments.

There are several advantages of Spearman's rho compared to the Bravais-Pearson correlation:
In many situations, e.g.\ if the data is non-elliptical, the Bravais-Pearson correlation may not be an appropriate measure for dependence.
It is confined to measuring linear dependence, while the rank-based dependence measure Spearman's rho quantifies monotone dependence.
If the second moments do not exist, the Bravais-Pearson correlation is not even defined, while Spearman's rho does not require any moments.

Spearman's rho is probably the most common rank-based dependence measure in economic and social sciences, see e.g.\ \citet{gaissler:2010}, who propose tests for equality of rank correlations, and the references herein.
In addition, Spearman's rho often performs better in terms of robustness than the Bravais-Pearson correlation.
\citet{embrechts:2002} discuss several other pitfalls and possible problems for a risk manager who simply applies the Bravais-Pearson correlation.

Therefore, it is natural in the context of testing for changes in
the dependence structure of random vectors to extend the
BPC test to a test for constant Spearman's rho. As
expected from the theory of dependence measures, this test is
applicable in more situations: It has a much better behavior in
the presence of outliers and there are no conditions on the
existence of moments. In addition,
the test is applicable in arbitrary dimensions, while the
BPC test is designed for bivariate random vectors.
Similarly to the BPC test, the test bases on successively calculated empirical correlation coefficients in the style of \citet{ploberger:1989}, \citet{lee:2003} or \citet{galeano:2007b}.

The limit distribution of our test statistic is the supremum of the absolute value of a Brownian bridge.
This immediately provides critical values without any bootstrap
techniques. We impose a strong mixing assumption for the
dependence structure.
The proof relies on an invariance principle for multivariate sequential empirical processes from \citet{buecher:2011}.

By using the copula-based expression for Spearman's rho from
\cite{schmid:2007} or \citet{nelsen:2006}, we get quite another
contribution with our test, i.e.\ an extension of the copula
constancy tests proposed by \citet{busetti:2011} and \citet{kraemervankampen:2011}. Since copula models are frequently used in financial econometrics (see e.g. \citealp{manner:2011} and \citealp{giacomini:2009}), such tests for structural change are important in this area. However, they are restricted to the case of testing for copula constancy in one parti\-cular quantile,
e.g.\ the $0.95$-quantile. This might be an important null
hypothesis as well, but our test now (indirectly) allows for testing constancy
of the whole copula by integrating over it.
We therefore reject the null hypothesis of constant Spearman's rho (which is closely connected to the null hypothesis of an overall constant copula) if the integral over it fluctuates too much over time.
The problem of testing constancy of the whole copula has recently been dealt with in the literature (see \citealp{kojadinovic:2012}, \citealp{ruppert:2011}, \citealp{VankampenWied:2012} and \citealp{remillard:2010}). All these approaches, however, need the computationally intensive bootstrap for approximating the limit distribution.
With our approach, we need much less computational time for calculating the critical values. \\
The paper is organized as follows: Section 2 presents our test
statistic and the asymptotic null distribution, Section 3
considers local power, Section 4 presents Monte Carlo evidence
about the behavior of the test in small samples. Section 5 compares
our new test with the BPC test in terms of
robustness by a simulation study and an empirical application.
Finally, Section 6 concludes. All proofs are in the appendix.

\section{Test statistic and its asymptotic null distribution}
In this section, we present the test statistic and the limit
distribution of our test under the null. First, we introduce some notation: $({\bf X}_1,\ldots,{\bf X}_n)$ are
$d$-dimensional random vectors on the probability space $(\Omega,\mathfrak{A},\mathsf{P})$ with ${\bf X}_j =
(X_{1,j},\ldots,X_{d,j}), j = 1,\ldots,n$. Regarding the dependence structure, we impose the following assumption:
\renewcommand{\Ann}{A}
\renewcommand{\ann}{}
\renewcommand{\AnnNumm}{(\Ann \arabic{*}\ann)}
\begin{enumerate}[label=\AnnNumm]
\item\label{alphamixing} ${\bf X}_1,\ldots,{\bf X}_n$ are $\alpha$-mixing with mixing coefficients $\alpha_j$ satisfying
\begin{align*}
\sum_{j=1}^{\infty} j^2 \alpha_j^{\gamma/(4+\gamma)} < \infty
\end{align*}
for some $\gamma \in (0,2)$.
\end{enumerate}
This dependence assumption is similar to the assumption made in \citet{inoue:2001} and holds in most econometric
models relevant in practice, e.g.\ for ARMA- and GARCH-processes under mild additional conditions, see e.g.\ \citet{carrasco:2002}. \\
The vectors ${\bf X}_j, j=1,\ldots,n$, have joint distribution
functions $F^j$ with
\begin{align*}
F^j({\bf x}) = \mathsf{P}(X_{1,j} \leq x_1,\ldots,X_{d,j} \leq x_d),
{\bf x} = (x_1,\ldots,x_d) \in \mathbb{R}^d,
\end{align*}
and marginal distribution functions $F_{i,j}(x) = \mathsf{P}(X_{i,j} \leq x)$ for $x \in \mathbb{R}$ and $i=1,\ldots,d$ which are assumed to be continuous. \\
According to Sklar's (1959)\nocite{sklar:1959} theorem, there exists a
unique copula function $C_j: [0,1]^d \rightarrow [0,1]$ of ${\bf X}_j$
with
\begin{align*}
F^j({\bf x}) = C_j(F_{1,j}(x_1),\ldots,F_{d,j}(x_d))
\end{align*}
and
\begin{align*}
C_j({\bf u}) = F_j(F_{1,j}^{-1}(u_1),\ldots,F_{d,j}^{-1}(u_d)),
{\bf u} = (u_1,\ldots,u_d) \in [0,1]^d,
\end{align*}
where $F^{-1}$ is the generalized inverse function, see e.g. \cite{schmid:2007}.

In terms of the copula, Spearman's rho is defined as
\begin{align*}
\rho_j = h(d) \cdot \left(2^d \int_{[0,1]^d} C_j({\bf u}) d {\bf
u} - 1 \right)
\end{align*}
with $h(d) = \frac{d+1}{2^d - (d+1)}$, see \cite{schmid:2007} or \citet{nelsen:2006}. There are also other possibilities to define Spearman's rho in higher dimensions (see \citealp{schmid:2007} and \citealp{quessy:2009}), but we focus on this expression for ease of exposition and because
this measure performs well in terms of the asymptotic relative efficiency compared with other multivariate extensions, see \citet[p. 328]{quessy:2009}. \\
We test
\begin{align*}
H_0: \rho_j = \rho_0, j=1,\ldots,n \text{ vs. } H_1: \exists j \in
\{1,\ldots,n-1\}: \rho_j \neq \rho_{j+1}.
\end{align*}
Let
\begin{align*}
\hat F_{i;n}(x) = \frac{1}{n} \sum_{j=1}^n {\bf 1}_{\{X_{i,j} \leq
x\}}, i=1,\ldots,d, x \in \mathbb{R},
\end{align*}
$U_{i,j} = F_{i,j}(X_{i,j})$ and
\begin{align*}
\hat U_{i,j;n} := \hat F_{i;n}(X_{i,j}) =
\frac{1}{n} \cdot \left(\text{rank of } X_{i,j} \text{ in }
X_{i,1},\ldots,X_{i,n}\right), i=1,\ldots,d,j=1,\ldots,n.
\end{align*}
Let $R_j({\bf u}) = {\bf 1}_{\{U_{1,j} \leq u_1,\ldots,U_{d,j} \leq u_d\}}$ and
$\hat R_j({\bf u}) = {\bf 1}_{\{\hat U_{1,j;n} \leq u_1,\ldots,\hat U_{d,j;n} \leq u_d\}}$.
The copula $C$ is estimated by the empirical copula, defined as
\begin{align*}
\hat C_n({\bf u}) = \frac{1}{n} \sum_{j=1}^n \hat R_j({\bf u}) =
\frac{1}{n} \sum_{j=1}^n \prod_{i=1}^d {\bf 1}_{\{\hat U_{i,j;n}
\leq u_i\}}, {\bf u} = (u_1,\ldots,u_d) \in [0,1]^d.
\end{align*}
The estimator based on the first $k$ observations is
\begin{align}\label{hatck}
\hat C_k({\bf u}) = \frac{1}{k} \sum_{j=1}^k \hat R_j({\bf u}) =
\frac{1}{k} \sum_{j=1}^k \prod_{i=1}^d {\bf 1}_{\{\hat U_{i,j;n}
\leq u_i\}}, {\bf u} = (u_1,\ldots,u_d) \in [0,1]^d.
\end{align}
Note that the application of the limit theorem from \citet{buecher:2011} requires that we use
$\hat U_{i,j;n}$ and not $\hat U_{i,j;k}$ in \eqref{hatck}. \\
The estimator for the copula immediately yields an estimator for
Spearman's rho:
\begin{align*}
\hat \rho_k = h(d) \cdot \left(2^d \int_{[0,1]^d} \hat C_k({\bf
u}) d {\bf u} -1 \right) = h(d) \cdot \left(\frac{2^d}{k}
\sum_{j=1}^k \prod_{i=1}^d (1- \hat U_{i,j;n}) -1 \right).
\end{align*}
We use the test statistic $W$, defined as
\begin{align*}
W = \hat D \max_{1 \leq k \leq n} \left|\frac{k}{\sqrt{n}} \left(\hat \rho_{k} - \hat \rho_n \right) \right| =: \hat D \sup_{s \in [0,1]} |P_n(s)|
\end{align*}
with $P_n(s) = \frac{[ns]}{\sqrt{n}} \left(\hat \rho_{[ns]} - \hat \rho_n \right)$ and with a deviation estimator $\hat D = \frac{1}{\sqrt{\hat D'}}$,
where
\begin{align*}
\hat D' &= h(d)^2 2^{2d} \left\{\frac{1}{n} \sum_{j=1}^n \prod_{i=1}^d (1-\hat U_{i,j;n})^2 - \left(\frac{1}{n} \sum_{j=1}^n \prod_{i=1}^d (1- \hat U_{i,j;n}) \right)^2 \right. \\ &\left. + 2 \left[\sum_{m=1}^{n-1} k\left(\frac{m}{\gamma_n}\right) \left(\sum_{j=1}^{n-m} \frac{1}{n} \prod_{i=1}^d (1-\hat U_{i,j;n})(1-\hat U_{i,j+m;n}) - \left(\frac{1}{n} \sum_{j=1}^n \prod_{i=1}^d (1- \hat U_{i,j;n}) \right)^2 \right) \right] \right\}.
\end{align*}
The kernel $k(\cdot)$ is selected from the class $\mathcal{K}_2$ of \citet{andrews:1991}, and the bandwidth $\gamma_n$ is chosen such that
$\gamma_n = o(n^{\frac{1}{2}})$, see Appendix A for details.
The estimator $\hat D$ is a scaling factor which takes serial dependence and fluctuations of the estimators for Spearman's rho into account.
The weighting factor $\frac{k}{\sqrt{n}}$ compensates for the fact that Spearman's rho can be estimated more precisely for larger $k$. \\
For our main theorem, we need two additional assumptions:
\renewcommand{\Ann}{B}
\renewcommand{\ann}{}
\renewcommand{\AnnNumm}{(\Ann \arabic{*}\ann)}
\begin{enumerate}[label=\AnnNumm]
\item\label{assump2a} $({\bf X}_1,\ldots,{\bf X}_n)$ is strictly stationary.
\item\label{assump2b} The copula $C$ and the marginal distribution functions $F_{i,j} = F_i, i=1,\ldots,d$ are continuous.
\end{enumerate}
Note that Assumption \ref{assump2a} is in line with other tests for structural change, see e.g. \citet{inoue:2001}.
\begin{theorem}\label{theorem1} Under $H_0$ and Assumptions \ref{alphamixing}, \ref{assump2a}, \ref{assump2b},
\begin{align*}
W \rightarrow_d \sup_{s \in [0,1]} |B(s)|,
\end{align*}
where $B(s)$ is a one-dimensional Brownian bridge.
\end{theorem}
Theorem \ref{theorem1} allows for constructing an asymptotic test. The main tool for the proof is an invariance principle from \citet{buecher:2011}
for the multivariate sequential rank order process
\begin{align*}
A_n(s,{\bf u}) &:= \frac{1}{\sqrt{n}} \sum_{j=1}^{[ns]} (\hat R_j({\bf u}) - \hat C_n({\bf u})) \\
               &= \frac{[ns]}{\sqrt{n}} \left(\frac{1}{[ns]}\sum_{j=1}^{[ns]} \hat R_j({\bf u}) - \hat C_n({\bf u}) \right),
\end{align*}
which was introduced by \citet{ruschendorf:1976}. This limit condition does not require any smoothness assumption on the derivatives of the copula, see \citet{segers:2011} for a detailed discussion of this issue. \\
There exists an interesting relationship between our test for
constancy of Spearman's rho and the copula constancy tests
proposed by \citet{busetti:2011} and by \citet{kraemervankampen:2011}:
One can show (see Appendix B) that (if $n \cdot u_i$ is an integer for all $i=1,\ldots,d$) our test is as a functional of
the multivariate $\tau$- (or $u$-)quantics on which these copula constancy
tests base. But, in fact, while the other tests examine if the
copula in a particular quantile is constant, we can test for
constancy of the whole copula by integrating over it. Although this is not the null hypothesis we consider, our test complements the literature on copula constancy tests.

Note that the integral of the copula (or more precisely, of the multivariate sequential rank order process) is a particular functional which has nice properties. For example, it leads to a simple limit distribution which is free of nuisance parameters, and therefore needs no bootstrap approximations.
The limit distributions are more involved and/or difficult to derive with other functionals such as used in \citet{kojadinovic:2012}, \citet{ruppert:2011}, \citet{VankampenWied:2012} and \citet{remillard:2010}.
(\citealp{VankampenWied:2012} use a slightly different definition of the multivariate sequential rank order process in that they consider the empirical quantile function $\hat F_n^{-1}$
and not the empirical distribution function $\hat F_n$.)

\section{Local power}
This section considers the local power of our test. Since the
copula function of the random vectors under consideration changes
with $n$, we now work with a triangular array $({\bf
X}_1^n,\ldots,{\bf X}_n^n)$, but we suppress the index $n$ for
ease of exposition. Let $C({\bf u})$ be a copula and let
$C^{'}(s,{\bf u})$ be another copula with an additional index
parameter $s$.  We consider local alternatives of the form
\begin{align}\label{wunschlocalpower}
C_j({\bf u}) = \left(1-\frac{\delta}{\sqrt{n}}\right) C({\bf u}) + \frac{\delta}{\sqrt{n}} C^{'}\left(\frac{j}{n},{\bf u}\right).
\end{align}
By choosing, say, $C^{'}\left(s,{\bf u}\right) = \left[1 - g(s)
\right] C\left({\bf u}\right) + g(s) C^{''}\left({\bf u}\right)$
for some copula $C^{''}(\cdot)$ and some function $g(\cdot)$
bounded by $1$ we obtain the sequence of correlations
\begin{align*}
\rho_j = \left[1-\frac{\delta}{\sqrt{n}} g\left(\frac{j}{n}\right)\right] \rho_0 + \frac{\delta}{\sqrt{n}} g\left(\frac{j}{n}\right) \rho_A.
\end{align*}
Choosing e.g. $g(s) = {\bf 1}_{\{s \geq 1/2\}}$ would lead to local alternatives in which the copula changes after the middle of the sample. A continuous function $g$ would lead to continuously changing copulas against which our test has power as well.

To deduce limit results for the sequence of local alternatives
\eqref{wunschlocalpower}, we need some more assumptions:
\renewcommand{\Ann}{C}
\renewcommand{\ann}{}
\renewcommand{\AnnNumm}{(\Ann \arabic{*}\ann)}
\begin{enumerate}[label=\AnnNumm]
\item\label{assump3a} The analogue of mixing condition \ref{alphamixing} holds for the triangular array.
\item\label{assump3e} The marginal distribution functions of $({\bf X}_1,\ldots,{\bf X}_n)$ do not depend on $j$ and are continuous.
\item\label{assump3b} The joint copula for the random vectors $({\bf X}_1,\ldots,{\bf
X}_n)$ with lag $l$,
\begin{align*}
C_{j,l}({\bf u_1},{\bf u_2}) :=\ &\mathsf{P}(X_{1,j} \leq F_{1,j}^{-1}(u_1^1),\ldots,X_{d,j} \leq F_{d,j}^{-1}(u_1^d), \\
&X_{1,j+l} \leq F_{1,j+l}^{-1}(u_2^1),\ldots,X_{d,j+l} \leq F_{d,j+l}^{-1}(u_2^d)),
\end{align*}
is specified to
\begin{align*}
C_{j,l}({\bf u_1},{\bf u_2}) &= \left(1 - \frac{\delta}{\sqrt{n}}\right)^2 C_l({\bf u_1},{\bf u_2}) + \frac{\delta^2}{n} C^{'}_l\left(\frac{j}{n},\frac{j+l}{n},{\bf u_1},{\bf u_2}\right) \\
&+ \frac{\delta}{\sqrt{n}} \left(1 - \frac{\delta}{\sqrt{n}}\right) \left[ C({\bf u_1}) C^{'}\left(\frac{j+l}{n},{\bf u_2}\right) + C({\bf u_2}) C^{'}\left(\frac{j}{n},{\bf u_1}\right) \right]
\end{align*}
with a constant $\delta \in (0,1]$. Here,
$C_l(\cdot,\cdot)$ is the joint copula of some sequence of
stationary random vectors $\xi_i$ with lag $l$, $C(\cdot)$ is the copula of $\xi_i$. Both $C_l(\cdot,cdot)$ and $C(\cdot)$ are assumed to be continuous. Analogously, $C^{'}_l(\cdot,\cdot,\cdot,\cdot)$
is the copula of some sequence of stationary random functions
$\eta_i(\cdot)$ with lag $l$, $C^{'}(\cdot,\cdot)$ is the copula of
$\eta_i(\cdot)$. The term $C'(r,t)$ is continuous in $t$ for all $r$ and there is $(r,r',t)$ such that $C'(r,t) \neq C'(r',t)$.
\end{enumerate}
The slightly cumbersome Assumption \ref{assump3b} is similar to Assumption B in \citet{inoue:2001} and is required for applying the limit theorem for the sequential empirical process under local alternatives and mixing conditions from \citet{inoue:2001}. Passing each element of ${\bf u_2}$ to $1$ yields equation \eqref{wunschlocalpower}.

With these assumptions, we get

\begin{theorem}\label{theorem2} Under Assumptions \ref{assump3a} and \ref{assump3b},
\begin{align*}
W \rightarrow_d \sup_{s \in [0,1]} \left| B(s) + \delta D h(d) 2^d
\left[\int_{[0,1]^d} \int_0^s C^*\left(t,{\bf u}\right) dt d {\bf
u} - s \int_{[0,1]^d} \int_0^1 C^*\left(t,{\bf u}\right) dt d {\bf
u} \right] \right|,
\end{align*}
where $D$ is the probability limit of $\hat D$ under the null
hypothesis.
\end{theorem}
With this theorem and Anderson's Lemma we can deduce that the
asymptotic level is always larger than or equal to $\alpha$, see
\citet{andrews:1997} or \citet{rothe:2011}.

\section{Finite sample behavior}\label{finite sample behavior}
We investigate the test's finite sample behavior and compare it to
the BPC test by simulating the empirical size under
the null hypothesis and the empirical power under various
alternatives in different settings. Some complementary simulation results are provided in \citet{dehling:2012} who propose a bivariate test for constant Kendall's tau.

In order to make the simulation study not too lengthy, we always use the Bartlett kernel
and bandwidth $[\log(n)]$ in the deviation estimator $\hat D$ of our test and the deviation estimator $\tilde D$ of the BPC test (which is described and denoted as $\hat D$ in \citealp{wied:2012}). Moreover, we restrict ourselves to the Student copula which is one of the most common copulas for modeling financial returns according to \citet[p. 181]{cherubini:2004}. These choices are made in \citet{wied:2012} as well. We consider three sample sizes $n=500,1000,2000$, the significance level $\alpha = 0.05$ and $50000$ repetitions in each setting. Moreover, we assume a bivariate $MA(1)$-process
\begin{align*}
{\bf X}_t = \epsilon_t + \theta \epsilon_{t-1} \text{ with }
\theta = \begin{pmatrix} \theta_1 & 0 \\ 0 & \theta_2
\end{pmatrix}, t = 1,\ldots,n,
\end{align*}
with, on the one hand, $(\theta_1,\theta_2)=(0,0)$ which corresponds to serial independence and, on the other hand, $(\theta_1,\theta_2)=(0.3,0.2)$ which corresponds to serial dependence. The $\epsilon_t, t = 0,1,\ldots,n$, are
independent and identically distributed, following a bivariate $t_{\nu}$-distribution with shape matrix
\begin{align} \label{matrix S}
S = \begin{pmatrix} 1 & q \\ q & 1 \end{pmatrix},\ |q| < 1.
\end{align}
We consider three different degrees of freedom, $\nu = 1,3,5$. In the case of $\nu = 1$, we do not have finite fourth moments (even no finite first moment),
which are required for the BPC test. Note that, in this case, the Bravais-Pearson correlation is not even defined. However, also in such a situation, it might be interesting to know if there is a structural change in the dependence structure.
The null hypothesis is that ${\bf X}_t$ has constant correlation
of $\rho_0=0.4$. Additionally, we consider seven alternatives, in which the correlation jumps after the middle of the sample
from $\rho_0=0.4$ to $\rho_1=0.6,0.8,0.2,0.0,-0.2,-0.4,-0.6$,
respectively. We simulate realizations
$\epsilon_0,\epsilon_1,\ldots,\epsilon_{n/2}$ and
$\epsilon_{n/2+1},\ldots,\epsilon_n$ with $q_i = \rho_i
\sqrt{\frac{(\theta_1^2 + 1)(\theta_2^2 + 1)}{\theta_1 \theta_2 +
1}}, i=0,1$. The choice of the $q_i$ is due to the fact
that with this, the Pearson correlation would be equal to $\rho_0$
resp. $\rho_1$ if it existed. Spearman's rho lies then closely to
these values as numerical approximations suggest.

Table \ref{simupowerdaniel2} reports rejection frequencies for serial independence and Table \ref{simupowerdaniel} for serial dependence.
\begin{center}
-Table \ref{simupowerdaniel2} here -

-Table \ref{simupowerdaniel} here -
\end{center}
At first, we discuss the case of $\nu = 1$. We see that the size of our test is kept and that the empirical power
increases with the magnitude of the break and sample size $n$. It is slightly higher for
increasing than for decreasing correlations. The BPC test is not applicable, because it cannot distinguish between the null and alternative hypothesis. This is partially due to the fact that the
asymptotic variance of the empirical correlation coefficient is an
unbounded function of the fourth moments of the population
distribution and that Spearman's correlation
coefficient is invariant under monotonely increasing,
componentwise transformations and hence
little effected by heavy tails.

For the $t_3$- and $t_5$-distribution, the power of our copula-based test is rather low compared to the BPC test. The efficiency becomes lower when the distribution comes ``closer'' to the Gaussian distribution. This is not surprising as for Gaussian data, the usual empirical correlation coefficient is the maximum likelihood, i.e.\ the most efficient estimator of the correlation. Note however that there is compelling empirical evidence that financial returns are not Gaussian distributed.

Interestingly, the power of our test is considerably lower (up to $15$ percentage points) in the case of serial dependence as compared to independence, which is not true for the BPC test. This indicates the potential drawback that the Spearman test is more sensitive with respect to serial dependence.

While the BPC test generally detects increasing correlations better than decreasing correlations, this does not hold for the Spearman test.

Repeating the experiments from Table \ref{simupowerdaniel} with the $t_2$-distribution under serial dependence and independence yields a better behavior for the BPC test as compared to the case of $\nu = 1$, but the test still does not keep its size.
Our test keeps its size, and its power is generally higher than the BPC power for decreasing correlations (especially for high shifts and large sample sizes). For increasing correlations, this is true for serial independence, large sample sizes and high shifts.\ Detailed results are available upon request.\ It is an interesting task for further research to discover the area of degrees of freedom and of serial dependence in which the application of the Spearman test can be recommended.

\begin{figure}
\centering
\includegraphics[width=\textwidth]{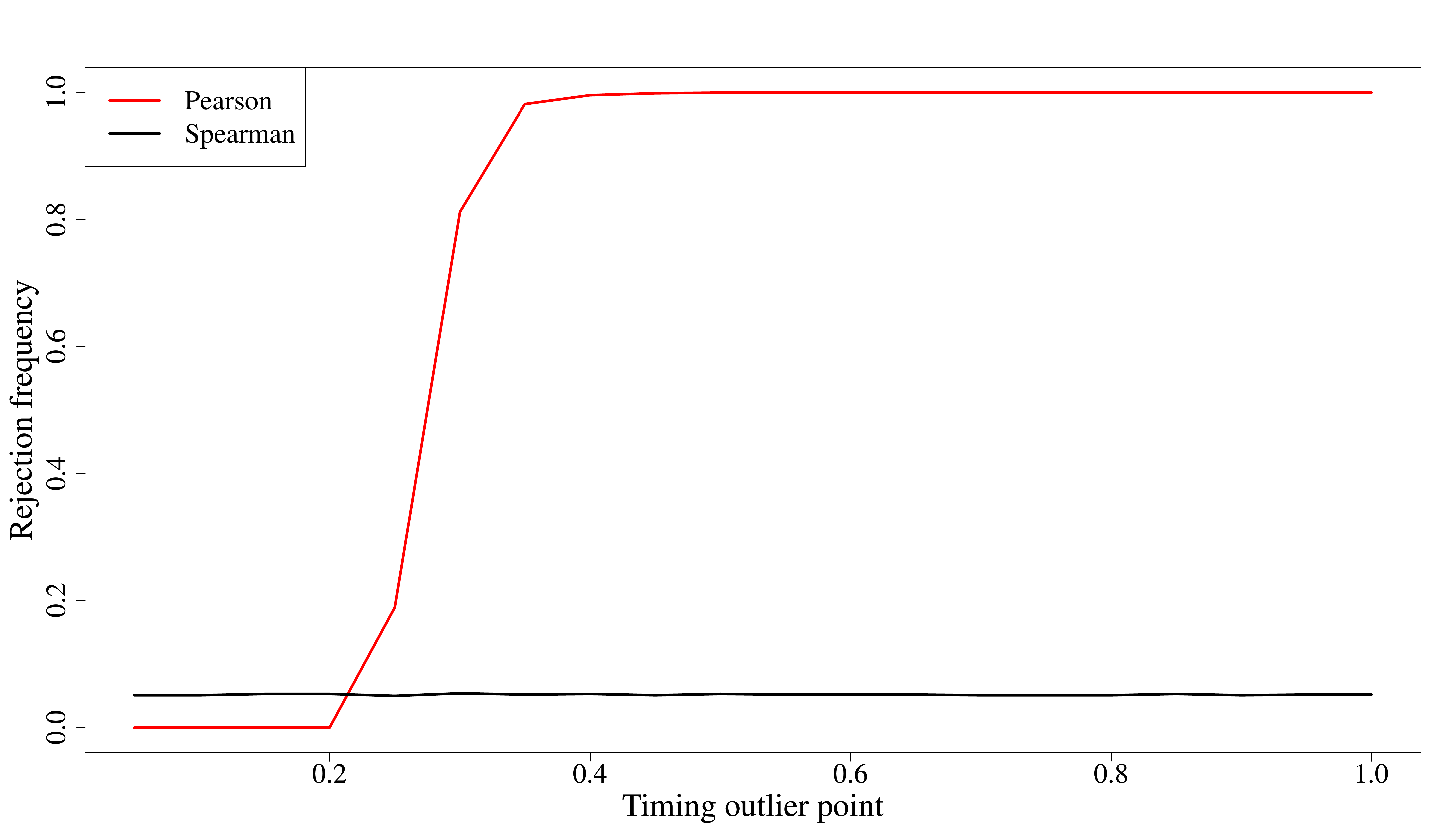}
\caption{Empirical rejection frequencies in an outlier scenario} \label{simupoweroutlier}
\end{figure}

Next, we consider an outlier scenario by using the setup from Table \ref{simupowerdaniel}
with the $t_5$-distribution, constant correlation $0.4$ and by
adding one heavy outlier of size, say,\\ $(40,-100)$ to the sample
at time $c \cdot 500, c=0.05,0.1,\ldots,1$. If
the outlier comes late, the BPC test almost always
rejects the null hypothesis; if it comes early, the test almost
never rejects it. This is an odd behavior and makes the test
unsuitable for this outlier scenario. Our new test always keeps
the size, see Figure \ref{simupoweroutlier}, whose message is basically the same in the presence of serial dependence.
\begin{center}
-Figure \ref{simupoweroutlier} here -
\end{center}
The power of our new test becomes higher when testing for a correlation change in more than two dimensions, and when there is an equal change in every component; see Table \ref{simupowerthreedim} for exemplary results for the trivariate $t_3$-distribution under serial independence. Here, it seems that decreasing correlations are better detected. Also in the trivariate case, the power decreases in the presence of serial dependence. With more than two dimensions, a direct comparison with the BPC test is not possible because this test is restricted to two dimensions.
\begin{center}
-Table \ref{simupowerthreedim} here -
\end{center}

\section{Robustness}
\subsection{Simulation evidence}

The two major advantages of our test compared to the BPC test proposed by \cite{wied:2012} are its applicability without any moment conditions at all (as compared to the existence of fourth moments for the BPC test) and its appealing robustness properties. The latter shall be visualized by an instructive example.
Both fluctuation tests proposed here mainly derive their robustness properties from the respective properties of the underlying correlation measure. The robustness properties of Spearman's rho, along with several other correlation estimators, are studied in detail in \cite{croux:2010}.

Similarly to the scenario of Figure \ref{simupoweroutlier}, we sample a path $(\bx_t)_{t = 1,\ldots,n}$ of length $n=500$ of the bivariate MA(1) process
 ${\bf X}_t = \epsilon_t + \theta \epsilon_{t-1}$,
where the $\epsilon_t$, $t = 0,\ldots,n$, are i.i.d.\ with a centered bivariate Gaussian distribution and covariance matrix $S$.
We choose
\begin{align*}
\theta = \begin{pmatrix} 0.3 & 0 \\ 0 & 0.2 \end{pmatrix}
\end{align*}
and $S$ as in equation \eqref{matrix S} with $q$ such that $\rho_0=0.4$.

\begin{figure}
\centering
\includegraphics[width=\textwidth]{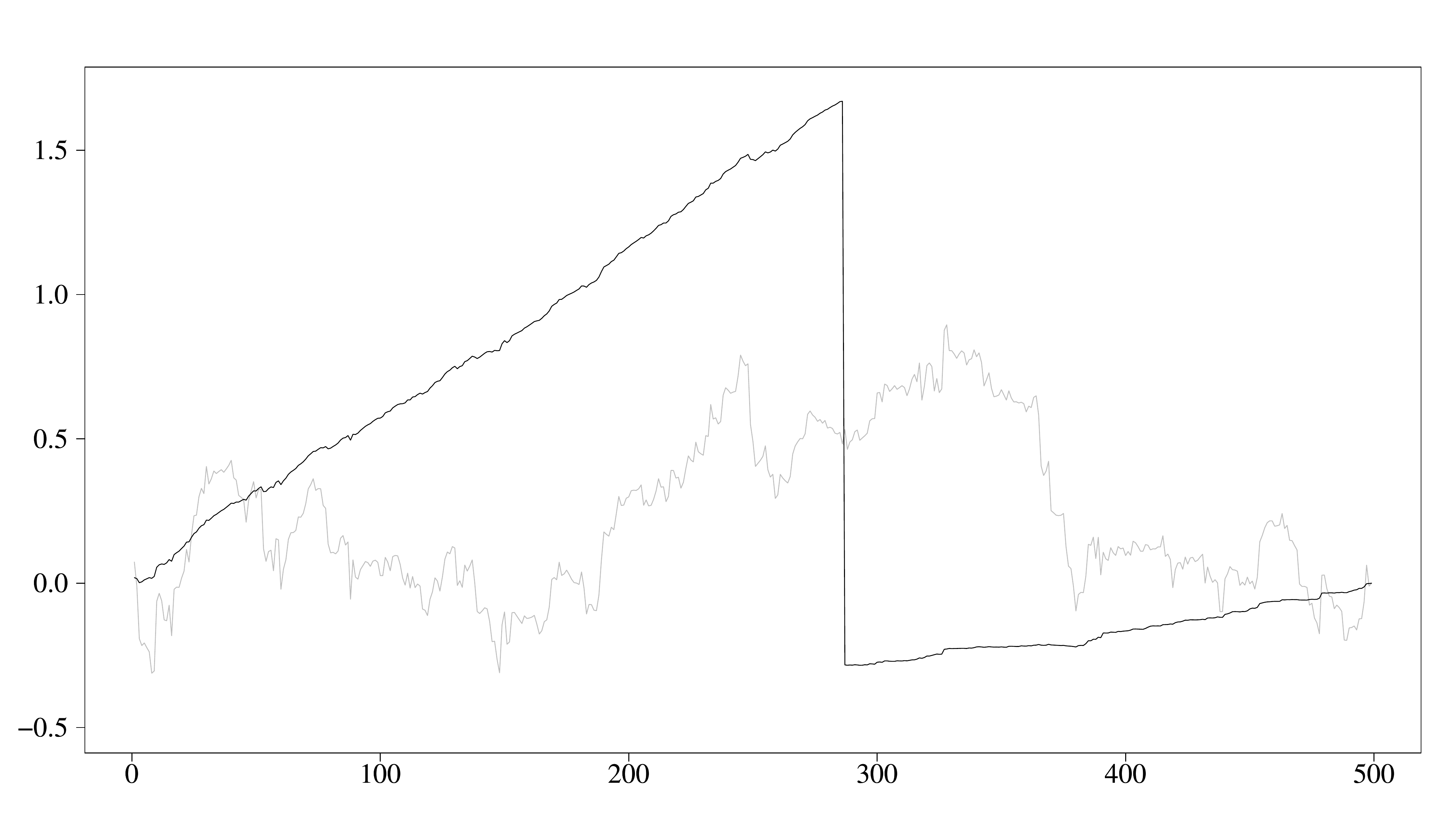}
\caption{Process $(b_k)_{k=1,...,500}$ computed from weakly contaminated (black) and uncontaminated (grey) sample} \label{Figure1}
\end{figure}

\begin{figure}
\centering
\includegraphics[width=\textwidth]{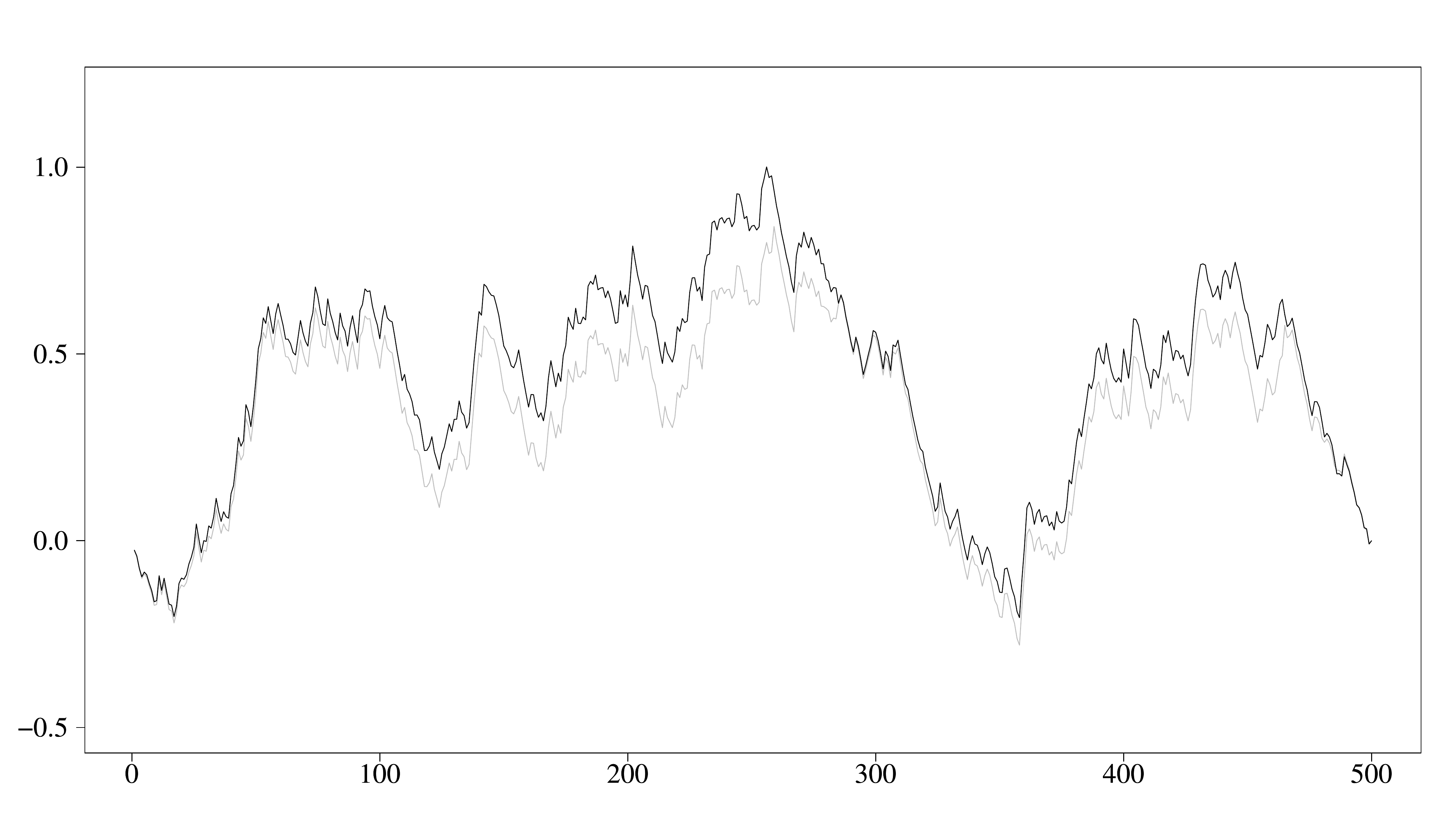}
\caption{Process $(\psi_k)_{k=1,...,500}$ computed from strongly contaminated (black) and uncontaminated (grey) sample} \label{Figure2}
\end{figure}

We add one mild outlier to the sample by setting, say, $\bx_{288}$ to $(20,-50)$.
We denote the resulting contaminated example by $(\bx_t^w)_{t = 1,...,n}$, where $w$ indicates \emph{w}eak contamination. Figure \ref{Figure1} visualizes the process
\begin{align*}
    b_k = \tilde{D} \frac{k}{\sqrt{n}} (\hat{r}_k - \hat{r}_n), \qquad k = 1,...,n,
\end{align*}
where $\hat{r}_k$ denotes the Bravais-Pearson correlation coefficient based on $\bX_1,...,\bX_k$, and where $\tilde{D}$ is the deviation estimator mentioned in the beginning of Section \ref{finite sample behavior} that scales the process such that $(b_{[ns]})_{s \in [0,1]}$ converges to a Brownian bridge.
\begin{center}
-Figure \ref{Figure1} here -
\end{center}
\begin{center}
-Figure \ref{Figure2} here -
\end{center}
The BPC test statistic is then $\sup_{1 \le k \le n} |b_k|$. The grey line in Figure \ref{Figure1} corresponds to the uncontaminated sample $(\bx_t)_{t=1,...,n}$, the black line to $(\bx_t^w)_{t=1,...,n}$. The single outlier has a dramatic effect on the Bravais-Pearson test statistic and, in this example, causes the null hypothesis to be rejected at the significance level $0.05$. While the position of the outlier influences the decision of the test (see Figure \ref{simupoweroutlier}), it has no big influence on the basic character of the shape of the black line.

Alternatively, we create a strongly contaminated sample $(\bx_t^s)_{t=1,...,n}$ by randomly placing $10$ outliers in the second half of the sample. Each outlier is of the form $(y_t,-y_t)$, where $y_t$ is drawn from the uniform distribution on $[-1000,-100] \cup [100,1000]$. Both, fraction and size of the outliers in $(\bx_t^s)_{t=1,...,n}$ are about 10 times as large as in $(\bx_t^w)_{t=1,...,n}$.
Figure \ref{Figure2} depicts the process
\begin{align*}
        \psi_k = \hat{D} \frac{k}{\sqrt{n}} (\hat{\rho}_k - \hat{\rho}_n), \qquad k = 1,...,n,
\end{align*}
once being computed from the uncontaminated sample (grey line) and once from the heavily corrupted sample $(\bx_t^s)_{t=1,...,n}$ (black line). We witness a slight distortion of $(\psi_k)_{k=1,...,n}$ as a result of the contamination, but the location of the maximizing point as well as the decision of the test are unaffected. Note that the main observations from Figure \ref{Figure1} and \ref{Figure2} basically could also be replicated under the $t_5$-distribution and/or serial indepedence as used in the previous section.

\subsection{Empirical relevance}
\begin{figure}[!h!]
\begin{center}
\subfigure[Process
$(b_k)_{k=1,\ldots,1262}$]{\includegraphics[scale=0.3]{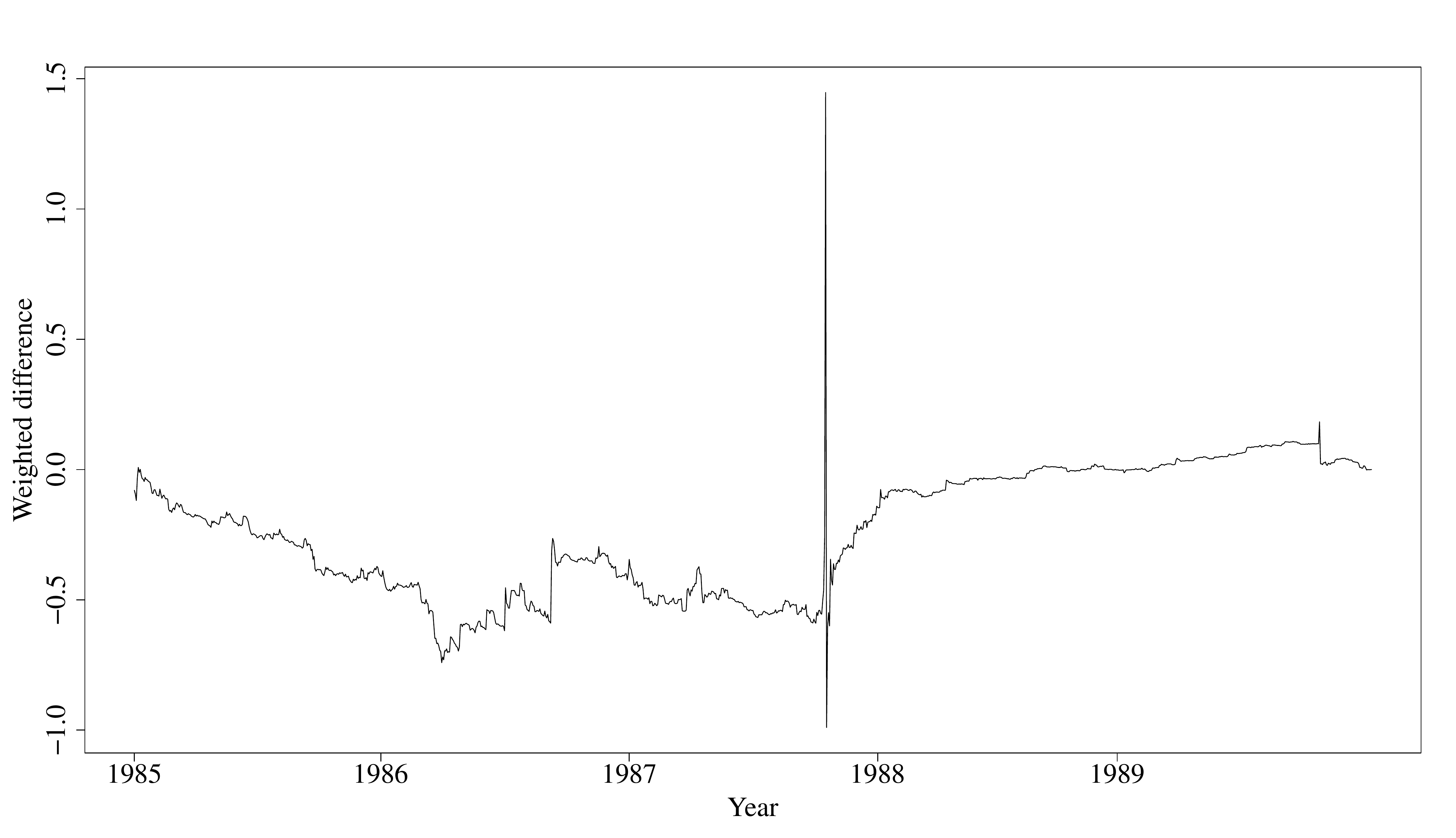}}
\subfigure[Process
$(\psi_k)_{k=1,\ldots,1262}$]{\includegraphics[scale=0.3]{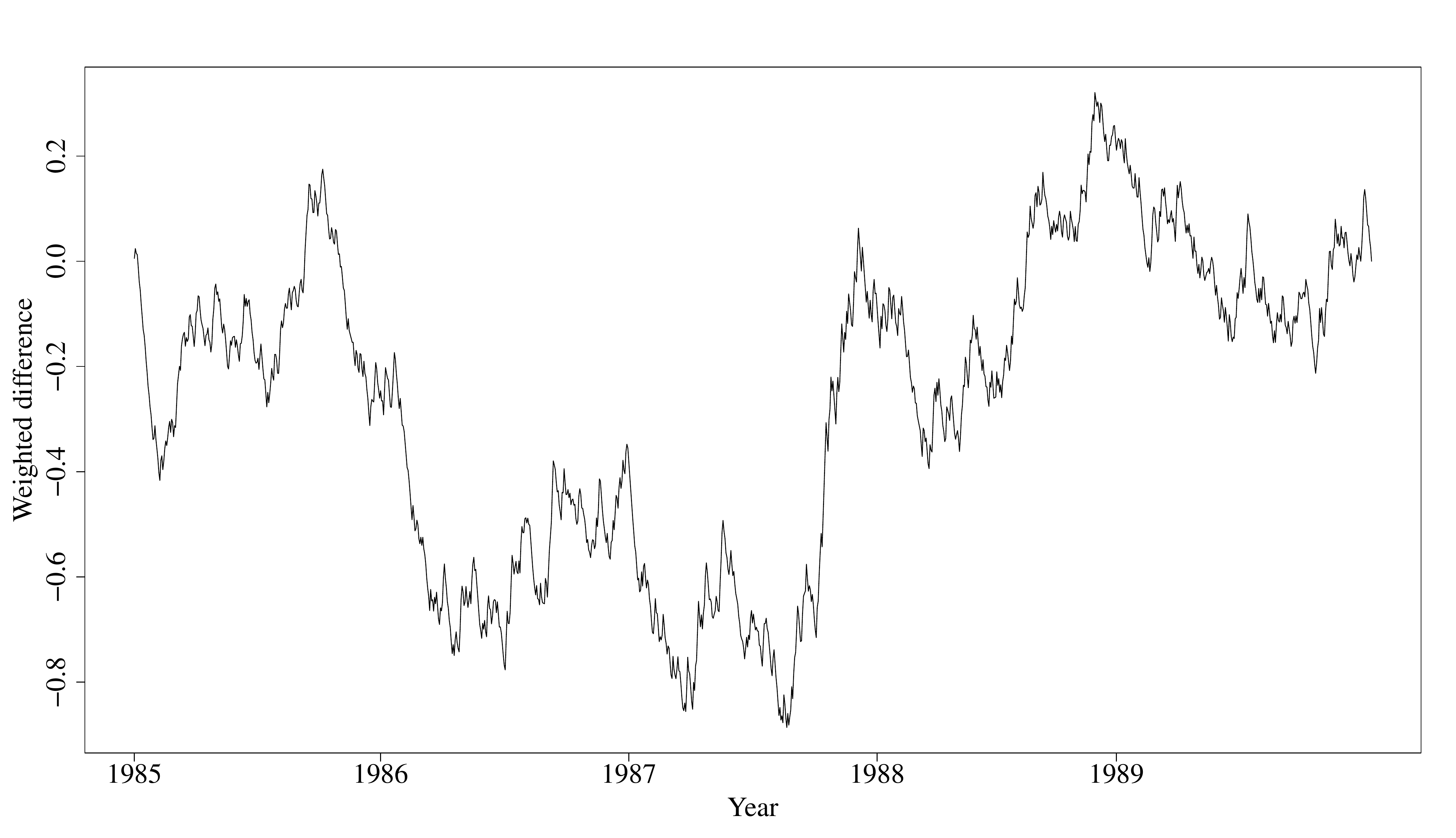}}
\caption{Processes $(b_k)_{k=1,\ldots,1262}$ and $(\psi_k)_{k=1,\ldots,1262}$ for the Dow Jones and Nasdaq
Index} \label{DowundNasdaq}
\end{center}
\end{figure}

This subsection shows that the outlier scenario described in the
previous subsection might indeed be relevant for a practitioner
who analyzes structural changes in the dependence of assets. This
can be exemplarily seen in the time period around the Black
Monday, 19th October 1987, i.e.\ in the time period from the beginning of January 1985 to the end of December 1989, using daily return data from Datastream (yielding $n=1262$). Both the Dow Jones Industrial Average and the
Nasdaq Composite daily returns are extremely negative at this day,
and the absolute values of these returns are much higher than the
other ones at this time. The next day, the Dow Jones return is
positive again, while the Nasdaq return remains negative - both on
a high level compared to the means and standard deviations of all
days. Table \ref{tablereturns} shows the exact values.
\begin{center}
-Table \ref{tablereturns} here -
\end{center}
These outliers are reflected in the BPC test statistic, see Figure \ref{DowundNasdaq}, part (a),
for a visualization of the process $(b_k)_{k=1,\ldots,1262}$ and the peak around the Black Monday. On
19th October, the successively estimated correlations become very
high, but fall down on the next day. Both phenomena together lead to the peak. Note that this is not exactly the same situation as in Figure \ref{Figure1}. However, by similar simulations with two outliers as in this application, we can reproduce the peak of Figure \ref{DowundNasdaq}, part (a), as well. Thus, the figures give two different examples of a bizarre behavior of the test statistic which are both due to the construction of the Pearson correlation coefficient. \\
Applying the BPC test gives a test statistic of $1.447$ ($p$-value of $0.030$) such that the null hypothesis of constant correlation is rejected at the $5 \%$-level. However, the test statistic would be much lower without this peak and the null would not be rejected. \\
Our Spearman test statistic is not affected by this peak - see
Figure \ref{DowundNasdaq}, part (b) for a visualization of the process $(\psi_k)_{k=1,\ldots,1262}$ - and the test statistic is
equal to $0.886$ ($p$-value of $0.412$). Therefore, one should
probably conclude that the dependence structure did not change
seriously after the Black Monday.
Similar results were obtained for other time
periods around 19th October 1987 and have been confirmed in \citet{dehling:2012} who perform a related analysis.
\begin{center}
-Figure \ref{DowundNasdaq} here -
\end{center}

\section{Discussion}
We propose a new test for constancy of Spearman's rho which is much more robust against outliers than the BPC test previously suggested by \citet{wied:2012}.

Indirectly, our test also allows for testing if the whole copula of
multivariate random vectors is constant, and thus it extends formerly suggested pointwise copula constancy tests. It is an interesting task for further research to compare the performance of our test with the performance of constancy tests for the whole copula, based on other functionals of the multivariate sequential rank order process, such as proposed in \citet{kojadinovic:2012}, \citet{ruppert:2011}, \citet{VankampenWied:2012} and \citet{remillard:2010}, especially in higher dimensions. Some limited evidence for the bivariate case (Spearman and maximum functional) can be found in \citet{VankampenWied:2012} and \citet{vankampen:2012}. In general, it is not obvious how to compare both tests as one needs to specify a bandwidth for the Spearman test and a block length for the maximum test. However, there is evidence that the Spearman test is typically (but not always) outperformed by the maximum test in terms of power. The Spearman test is computationally less intensive as no bootstrap approximations are required.

Another worthwile research approach would be an extension of the considered dependence structure to functionals of iid- or even of
mixing processes in order to enlarge the class of models in which our test
can operate.

\appendix

\section{Appendix section}
{\it Proof of Theorem \ref{theorem1}} \\
Denote with $l^{\infty}(\mathbb{R}^{k})$ the function space of all bounded functions from $\mathbb{R}^{k}$ to $\mathbb{R}$.

Consider first $P_n(s)$:
\begin{align*}
P_n(s) &= \frac{[ns]}{\sqrt{n}} \left(h(d) \cdot \left(2^d \int_{[0,1]^d} \hat C_{[ns]}({\bf u}) d {\bf u} -1 \right) - h(d) \cdot \left(2^d \int_{[0,1]^d} \hat C_n({\bf u}) d {\bf u} -1 \right) \right) \\
     &= \frac{[ns]}{\sqrt{n}} \cdot h(d) \cdot 2^d \int_{[0,1]^d} \left(\frac{1}{[ns]} \sum_{j=1}^{[ns]} \hat R_j({\bf u}) - \frac{1}{n} \sum_{j=1}^n \hat R_j({\bf u}) \right) d {\bf u} \\
     &= h(d) \cdot 2^d \int_{[0,1]^d} A_n(s,{\bf u}) d {\bf u}
\end{align*}
with
\begin{align}\label{an}
A_n(s,{\bf u}) = \frac{[ns]}{\sqrt{n}} \left(\frac{1}{[ns]} \sum_{j=1}^{[ns]} \hat R_j({\bf u}) - \frac{1}{n} \sum_{j=1}^n \hat R_j({\bf u}) \right).
\end{align}
With an invariance principle for
\begin{align}\label{invarianzprinzip}
\frac{1}{\sqrt{n}} \sum_{j=1}^{[ns]} (R_j({\bf u}) - C({\bf u})),
\end{align}
which is presented in Theorem 2.1 in \cite{inoue:2001} (setting $\delta$ from \citealp{inoue:2001} to $0$ and defining $x_{ni}$ from \citealp{inoue:2001} as $(F_i(X_{i,j}))_{1 \leq i \leq d}$), Condition 3.1 in \citet{buecher:2011} is satisfied with $G_n^*(s,{\bf u}) := \frac{1}{n} \sum_{j=1}^{[ns]} R_j({\bf u})$ and
$C^*(s,{\bf u}) := s C({\bf u})$ which lies in the space $\mathbb{D}_{\Psi}$ defined in \citet{buecher:2011}. So we obtain with
Corollary 3.3.a in \citet{buecher:2011}
\begin{align*}
A_n(\cdot,\cdot) \rightarrow_d A_0(\cdot,\cdot) \text{ in } l^{\infty}(\mathbb{R}^{d+1})
\end{align*}
with
\begin{align*}
A_0(s,{\bf u}) = V_0(s,{\bf u}) - s V_0(1,{\bf u}).
\end{align*}
$V_0(s,{\bf u})$ is a $\mathsf{P}$-almost surely continuous, centered Gaussian process with covariance function
\begin{align*}
K_0((s_1,{\bf u_1}),(s_2,{\bf u_2})) := \mathsf{Cov}(V_0(s_1,{\bf u_1}),V_0(s_2,{\bf u_2})) = (s_1 \wedge s_2)K'({\bf u_1},{\bf u_2}),
\end{align*}
where
\begin{align*}
K'({\bf u_1},{\bf u_2}) &= C({\bf u_1} \wedge {\bf u_2}) - C({\bf u_1}) C({\bf u_2}) \\
&+ \sum_{m=2}^{\infty} \left(\mathsf{E}({\bf 1}_{\{X_{1,1} \leq F_1^{-1}(u_1^1);\ldots;X_{d,1} \leq F_d^{-1}(u_1^d)\}} {\bf 1}_{\{X_{1,m} \leq F_1^{-1}(u_2^1);\ldots;X_{d,m} \leq F_d^{-1}(u_2^d)\}}) \right. \\
&\left. - \mathsf{E}({\bf 1}_{\{X_{1,1} \leq F_1^{-1}(u_1^1);\ldots;X_{d,1} \leq F_d^{-1}(u_1^d)\}}) \cdot \mathsf{E}({\bf 1}_{\{X_{1,m} \leq F_1^{-1}(u_2^1);\ldots;X_{d,m} \leq F_d^{-1}(u_2^d)\}}) \right) \\
&+ \sum_{m=2}^{\infty} \left( \mathsf{E}({\bf 1}_{\{X_{1,1} \leq F_1^{-1}(u_2^1);\ldots;X_{d,1} \leq F_d^{-1}(u_2^d)\}} {\bf 1}_{\{X_{1,m} \leq F_1^{-1}(u_1^1);\ldots;X_{d,m} \leq F_d^{-1}(u_1^d)\}}) \right. \\
&\left. - \mathsf{E}({\bf 1}_{\{X_{1,1} \leq F_1^{-1}(u_2^1);\ldots;X_{d,1} \leq F_d^{-1}(u_2^d)\}}) \cdot \mathsf{E}({\bf 1}_{\{X_{1,m} \leq F_1^{-1}(u_1^1);\ldots;X_{d,m} \leq F_d^{-1}(u_1^d)\}}) \right).
\end{align*}
This covariance function is the limit of the covariance function of
\begin{align*}
V_n(s,{\bf u}) &= \frac{1}{\sqrt{n}} \sum_{j=1}^{[ns]} (R_j({\bf u}) - C({\bf u})),
\end{align*}
i.e.\
\begin{align*}
K_0((s_1,{\bf u_1}),(s_2,{\bf u_2})) = \lim_{n \rightarrow \infty} \mathsf{Cov}(V_n(s_1,{\bf u_1}),V_n(s_2,{\bf u_2})),
\end{align*}
see \citet{inoue:2001}.
Now, with the continuous mapping theorem,
\begin{align*}
P_n(\cdot) \rightarrow_d P_0(\cdot) \text{ in } l^{\infty}(\mathbb{R}),
\end{align*}
where
\begin{align*}
P_0(s) = h(d) 2^d \int_{[0,1]^d} A_0(s,{\bf u}) d {\bf u}
\end{align*}
is a $\mathsf{P}$-almost surely continuous, centered Gaussian process.
With Fubini's theorem, the covariance function is
\begin{align*}
&\mathsf{Cov}(P_0(s_1),P_0(s_2)) \\ &= h(d)^2 2^{2d} \int_{[0,1]^d} \int_{[0,1]^d} \mathsf{Cov}(V_0(s_1,{\bf u_1}) - s_1 V_0(1,{\bf u_1}), V_0(s_2,{\bf u_2}) - s_2 V_0(1,{\bf u_2})) d {\bf u_1} d {\bf u_2} \\
&= h(d)^2 2^{2d} (s_1 \wedge s_2 - s_1 s_2 - s_1 s_2 + s_1 s_2) \int_{[0,1]^d} \int_{[0,1]^d} \mathsf{Cov}(V_0(1,{\bf u_1}),V_0(1,{\bf u_2})) d {\bf u_1} d {\bf u_2} \\
&= (s_1 \wedge s_2 - s_1 s_2) D'
\end{align*}
with
\begin{align*}
D' &= h(d)^2 2^{2d} \int_{[0,1]^d} \int_{[0,1]^d} K'({\bf u_1},{\bf
u_2}) d {\bf u_1} d {\bf u_2} \\
&= h(d)^2 2^{2d} \left\{\mathsf{E}\left(
\prod_{i=1}^d (1- U_{i,j})^2 \right) -
\left[\mathsf{E}\left(\prod_{i=1}^d (1- U_{i,j}) \right) \right]^2
\right. \\ &\left. + 2 \left[\sum_{m=1}^{\infty}
\mathsf{E}\left(\prod_{i=1}^d (1- U_{i,j})(1- U_{i,j+m}) \right)
- \left(\mathsf{E}\left(\prod_{i=1}^d (1- U_{i,j}) \right)
\right)^2 \right] \right\} \\ &= h(d) 2^{2d}
\left[\mathsf{Var}\left(\prod_{i=1}^d (1- U_{i,j})\right) +
2\sum_{m=1}^{\infty}\mathsf{Cov}\left(\prod_{i=1}^d (1-
U_{i,j}),\prod_{i=1}^d (1- U_{i,j+m})\right)\right].
\end{align*}
This holds because, again with Fubini,
\begin{align*}
&\int_{[0,1]^d} \int_{[0,1]^d} C({\bf u_1} \wedge {\bf u_2}) d {\bf u_1} d {\bf u_2} \\ &= \int_{[0,1]^d} \int_{[0,1]^d} \mathsf{E}({\bf 1}_{\{X_{1,j} \leq F_1^{-1}(u_1^1);\ldots;X_{d,j} \leq F_d^{-1}(u_1^d)\}} {\bf 1}_{\{X_{1,j} \leq F_1^{-1}(u_2^1);\ldots;X_{d,j} \leq F_d^{-1}(u_2^d)\}}) d {\bf u_1} d {\bf u_2} \\ &=
\mathsf{E} \left(\int_{[0,1]^d} \int_{[0,1]^d} {\bf 1}_{\{X_{1,j} \leq F_1^{-1}(u_1^1);\ldots;X_{d,j} \leq F_d^{-1}(u_1^d)\}} {\bf 1}_{\{X_{1,j} \leq F_1^{-1}(u_2^1);\ldots;X_{d,j} \leq F_d^{-1}(u_2^d)\}} d {\bf u_1} d {\bf u_2} \right) \\ &=
\mathsf{E} \left(\int_{[0,1]^d} \int_{[0,1]^d} {\bf 1}_{\{U_{1,j} \leq u_1^1;\ldots;U_{d,j} \leq u_1^d\}} {\bf 1}_{\{U_{1,j} \leq u_2^1;\ldots;U_{d,j} \leq u_2^d\}} d {\bf u_1} d {\bf u_2} \right) \\ &=
\mathsf{E} \left( \prod_{i=1}^d (1-U_{i,j}) \prod_{i=1}^d (1-U_{i,j}) \right).
\end{align*}
The other summands of $K'({\bf u_1},{\bf u_2})$ are integrated analoguesly. \\
We get a consistent estimator for $D'$ from \cite{davidson:2000},
\begin{align*}
\tilde D' &= h(d)^2 2^{2d} \left\{\frac{1}{n} \sum_{j=1}^n
\prod_{i=1}^d (1-U_{i,j})^2 - \left(\frac{1}{n}
\sum_{j=1}^n \prod_{i=1}^d (1- U_{i,j}) \right)^2 \right.
\\ &\left. + 2 \left[\sum_{m=1}^{n-1}
k\left(\frac{m}{\gamma_n}\right) \left(\sum_{j=1}^{n-m}
\frac{1}{n} \prod_{i=1}^d (1- U_{i,j})(1- U_{i,j+m})
- \left(\frac{1}{n} \sum_{j=1}^n \prod_{i=1}^d (1- U_{i,j})
\right)^2 \right) \right] \right\},
\end{align*}
with a kernel $k$ that is contained in the class $\mathcal{K}_2$ of \citet{andrews:1991} which guarantees positive semi-definiteness of $\tilde D'$.
Next, we show that $\hat D'- \tilde D' \rightarrow_p 0$. By the invariance principle \eqref{invarianzprinzip} (this time applied on the components of ${\bf X}_j$) we get a Glivenko-Cantelli-like theorem (in probability) with rate $n^{-\frac{1}{2}}$ for the marginal empirical distribution functions, that means,
\begin{align*}
 B_n:=\max_{i=1,\ldots,d} \sup_{j \in \mathbb{N}} |\hat{U}_{i,j;n}-U_{i,j}|
 =O_{\mathsf{P}}\left(n^{-\frac{1}{2}} \right).
\end{align*}
Since $0\leq U_{i,j}, \hat{U}_{i,j;n} \leq 1$, we obtain
\begin{align*}
  \left| \prod_{i=1}^d (1-\hat{U}_{i,j;n})(1-\hat{U}_{i,j+m;n})
  -\prod_{i=1}^d (1-U_{i,j})(1-U_{i,j+m})\right|  \leq 2d B_n.
\end{align*}
Thus we get
\begin{align*}
 |\hat D' -\tilde D'|\leq C \sum_{m=1}^{n-1}
 k\left(\frac{m}{\gamma_n}\right) B_n
 =O_{\mathsf{P}}\left( \gamma_n n^{-\frac{1}{2}} \right)=o_{\mathsf{P}}(1),
\end{align*}
as $\gamma_n=o(n^{\frac{1}{2}})$, compare the argument in \citet[p. 852]{andrews:1991}. Therefore $\hat D'$ is a
consistent estimator of $D'$. \\
The theorem follows then with the continuous mapping theorem,
because the process
\begin{align*}
P^*_0(s) := \frac{1}{\sqrt{D'}} P_0(s)
\end{align*}
is a $\mathsf{P}$-almost surely continuous, centered Gaussian process with the same covariance function as the Brownian bridge, i.e.\
\begin{align*}
\mathsf{Cov}(P^*_0(s_1),P^*_0(s_2)) = s_1 \wedge s_2 - s_1 s_2.
\end{align*}
Since a Gaussian process is uniquely determined by the first two moments, the limit process is in fact a Brownian bridge. \hfill $\blacksquare$ \\ \\

{\it Proof of Theorem \ref{theorem2}} \\
The proof is basically similar to the proof of Theorem \ref{theorem1}. We consider $A_n(s,{\bf u})$ from \eqref{an} and use a minor modification of Corollary 3.3.a in \citet{buecher:2011}.
Condition 3.1 is modified in that sense that we formulate the condition for triangular arrays. The invariance principle from Theorem 2.1 in \cite{inoue:2001}
(defining again $x_{ni}$ from \citealp{inoue:2001} as $(F_i(X_{i,j}))_{1 \leq i \leq d}$ with the stationarity of the marginal distribution functions) yields that
\begin{align}
\frac{1}{\sqrt{n}} \sum_{j=1}^{[ns]} (R_j({\bf u}) - C_j({\bf u}))
\end{align}
converges in distribution to the process $V_0(s,{\bf u})$ such that
\begin{align}
\frac{1}{\sqrt{n}} \sum_{j=1}^{[ns]} (R_j({\bf u}) - C({\bf u})) &= \frac{1}{\sqrt{n}} \sum_{j=1}^{[ns]} (R_j({\bf u}) - C_j({\bf u})) + \frac{1}{\sqrt{n}} \sum_{j=1}^{[ns]} (C_j({\bf u}) - C({\bf u})) \\
&= \frac{1}{\sqrt{n}} \sum_{j=1}^{[ns]} (R_j({\bf u}) - C_j({\bf u})) + \frac{\delta}{n} \sum_{j=1}^{[ns]} \left(C^{'}\left(\frac{j}{n},{\bf u}\right)  - C({\bf u}) \right)
\end{align}
converges in distribution to the process $V_0(s,{\bf u}) + \delta \int_0^s C^{'}\left(t,{\bf u}\right) dt - \delta s C({\bf u})$. This process fulfills Condition 3.1 in \citet{buecher:2011} with $G_n^*(s,{\bf u}) = \frac{1}{n} \sum_{j=1}^{[ns]} R_j({\bf u})$
and $C^*(s,{\bf u}) = s C({\bf u})$ which lies in the space $\mathbb{D}_{\Psi}$ defined in \citet{buecher:2011}. Thus the proof of Corollary 3.3.a goes through in the same way as in the stationarity case. Thus, $A_n(s,{\bf u})$ converges to
\begin{align*}
A_0(s,{\bf u}) + \delta \left[\int_0^s C^{'}\left(t,{\bf u}\right) dt - s \int_0^1 C^{'}\left(t,{\bf u}\right) dt \right].
\end{align*}
In addition, the probability limit of $\hat D$ under the sequence
of local alternatives is the quantity $D$ from the proof of
Theorem \ref{theorem1}, i.e.\ the probability limit of $\hat D$
under the null hypothesis. This holds because
\begin{align*}
\lim_{n \rightarrow \infty} \mathsf{Cov}(V_n(s_1,{\bf
u_1}),V_n(s_2,{\bf u_2}))
\end{align*}
is the same under the null hypothesis as well as under the
sequence of local alternatives.
Thus, the theorem is proved. \hfill $\blacksquare$ \\ \\

\section{Connection to copula constancy tests}
Let $\hat F_{i,n}^{-1}(u_i), 0 < u_i < 1, i=1,\ldots,d,$ denote the empirical quantile function and consider the case that $n \cdot u_i$ is an integer for all $i=1,\ldots,d$. Write
\begin{eqnarray*}
    A_n(s,u) &=& \frac{[ns]}{\sqrt{n}}\left(\frac{1}{[ns]}\sum_{j=1}^{[ns]}\hat R_j(u) - \frac{1}{n}\sum_{j=1}^{n}\hat R_j(u)\right) \\
&=& -\frac{[ns]}{\sqrt{n}}\left(\frac{1}{[ns]}\sum_{j=1}^{[ns]}\left(\hat C_n(u) - \hat R_j(u)\right)\right) + \frac{[ns]}{\sqrt{n}}\left(\frac{1}{n}\sum_{j=1}^{n}\left(\hat C_n(u) - \hat R_j(u)\right)\right) \\
&=& -\frac{1}{\sqrt{n}}\sum_{j=1}^{[ns]}\left(\hat C_n(u) - \hat R_j(u)\right) + \frac{1}{\sqrt{n}}\sum_{k=1}^{[ns]}\left(\frac{1}{n}\sum_{j=1}^{n}\left(\hat C_n(u) - \hat R_j(u)\right)\right) \\
&=&  -\frac{1}{\sqrt{n}}\sum_{j=1}^{[ns]} \left[\left(\hat C_n(u) - \hat R_j(u)\right) - \frac{1}{n}\sum_{m=1}^n \left(\hat C_n(u) - \hat R_m(u)\right)\right] \\
&=& -\frac{1}{\sqrt{n}}\sum_{j=1}^{[ns]} \left[\hat C_n(u) - {\bf 1}_{\{X_{1j} \leq \hat F_{1,n}^{-1}(u_1),\ldots, X_{dj}  \leq \hat F_{d,n}^{-1}(u_d)\}}\right],
\end{eqnarray*}
where the last step uses that $1/n\sum_{m=1}^n \hat R_m(u) = \hat C_n(u)$ and
\begin{eqnarray*}
  \hat R_j(u) &:=& {\bf 1}_{\{\hat U_{1j;n} \leq u_1,\ldots,\hat U_{dj;n} \leq u_d\}} \\
     &=& {\bf 1}_{\{\hat F_{1,n}^{-1}(\hat U_{1j;n}) \leq \hat F_{1,n}^{-1}(u_1),\ldots,\hat F_{d,n}^{-1}(\hat U_{dj;n})  \leq \hat F_{d,n}^{-1}(u_d)\}} \\
     &=& {\bf 1}_{\{X_{1j} \leq \hat F_{1,n}^{-1}(u_1),\ldots, X_{dj}  \leq \hat F_{d,n}^{-1}(u_d)\}}.
\end{eqnarray*}
Note that $\hat C_n(u) - {\bf 1}_{\{X_{1j} \leq \hat F_{1,n}^{-1}(u_1),\ldots, X_{dj}  \leq \hat F_{d,n}^{-1}(u_d)\}}$ are the bivariate $\tau$- or $u$-quantics in \citet{busetti:2011} if $d=2$.
Hence, our test bases on similar quantities. \\

{\it Acknowledgements:} Financial support by Deutsche
Forschungsgemeinschaft (SFB 823, Statistik nichtlinearer
dynamischer Prozesse, projects A1 and C3) and Ruhr Graduate School in Economics is gratefully acknowledged. We are grateful to helpful comments from the Editor, the Associate Editor and the referees. \\

\bibliography{spearman}

\newpage


\begin{table}[!h!]
\centering
\caption{Empirical power in different settings with serial independence, $\rho_0=0.4$, results of the Pearson-test in brackets\label{simupowerdaniel2}}
\vspace{0.5cm}
\begin{tabular}{ccccccccc}\hline \hline
 $n$ & \multicolumn{8}{c}{Values of $\rho_1$} \\
  & $0.4$ & $0.6$ & $0.8$ & $0.2$ & $0$ & $-0.2$ & $-0.4$ & $-0.6$ \\ \hline
             \multicolumn{8}{c}{a) bivariate $t_1$ distribution}\\
  $500$ & $0.045$ & $0.081$ & $0.205$ & $0.079$ & $0.199$ & $0.418$ & $0.702$ & $0.917$ \\
        & $(0.478)$ & $(0.507)$ & $(0.550)$ & $(0.474)$ & $(0.476)$ & $(0.490)$ & $(0.513)$ & $(0.533)$ \\
  $1000$ & $0.046$ & $0.121$ & $0.383$ & $0.120$ & $0.370$ & $0.730$ & $0.953$ & $0.998$ \\
        & $(0.477)$ & $(0.505)$ & $(0.553)$ & $(0.469)$ & $(0.479)$ & $(0.490)$ & $(0.511)$ & $(0.539)$ \\
  $2000$ & $0.048$ & $0.205$ & $0.662$ & $0.204$ & $0.656$ & $0.960$ & $1.000$ & $1.000$ \\
        & $(0.477)$ & $(0.500)$ & $(0.550)$ & $(0.471)$ & $(0.475)$ & $(0.489)$ & $(0.514)$ & $(0.538)$ \\ \hline
                 \multicolumn{8}{c}{b) bivariate $t_3$ distribution}\\
  $500$ & $0.045$ & $0.085$ & $0.217$ & $0.088$ & $0.230$ & $0.495$ & $0.786$ & $0.958$ \\
        & $(0.050)$ & $(0.260)$ & $(0.725)$ & $(0.125)$ & $(0.412)$ & $(0.699)$ & $(0.843)$ & $(0.907)$ \\
  $1000$ & $0.045$ & $0.129$ & $0.404$ & $0.134$ & $0.426$ & $0.804$ & $0.978$ & $1.000$ \\
        & $(0.040)$ & $(0.352)$ & $(0.845)$ & $(0.195)$ & $(0.602)$ & $(0.831)$ & $(0.916)$ & $(0.950)$ \\
  $2000$ & $0.048$ & $0.222$ & $0.699$ & $0.230$ & $0.732$ & $0.983$ & $1.000$ & $1.000$ \\
        & $(0.035)$ & $(0.479)$ & $(0.910)$ & $(0.310)$ & $(0.756)$ & $(0.906)$ & $(0.951)$ & $(0.968)$ \\ \hline
               \multicolumn{8}{c}{c) bivariate $t_5$ distribution}\\
  $500$ & $0.047$ & $0.085$ & $0.215$ & $0.089$ & $0.240$ & $0.515$ & $0.804$ & $0.966$ \\
        & $(0.038)$ & $(0.435)$ & $(0.958)$ & $(0.261)$ & $(0.778)$ & $(0.962)$ & $(0.990)$ & $(0.995)$ \\
  $1000$ & $0.047$ & $0.129$ & $0.404$ & $0.136$ & $0.446$ & $0.824$ & $0.984$ & $1.000$ \\
        & $(0.038)$ & $(0.693)$ & $(0.992)$ & $(0.506)$ & $(0.955)$ & $(0.993)$ & $(0.998)$ & $(0.999)$ \\
  $2000$ & $0.048$ & $0.226$ & $0.698$ & $0.237$ & $0.753$ & $0.987$ & $1.000$ & $1.000$ \\
        & $(0.037)$ & $(0.913)$ & $(0.998)$ & $(0.806)$ & $(0.992)$ & $(0.998)$ & $(0.999)$ & $(1.000)$ \\ \hline
        \end{tabular}
\end{table}

\begin{table}[!h!]
\centering
\caption{Empirical power in different settings with serial dependence, $\rho_0=0.4$, results of the Pearson-test in brackets
\label{simupowerdaniel}}
\vspace{0.5cm}
\begin{tabular}{ccccccccc}\hline \hline
 $n$ & \multicolumn{8}{c}{Values of $\rho_1$} \\
  & $0.4$ & $0.6$ & $0.8$ & $0.2$ & $0$ & $-0.2$ & $-0.4$ & $-0.6$ \\ \hline
             \multicolumn{8}{c}{a) bivariate $t_1$ distribution}\\
  $500$ & $0.053$ & $0.077$ & $0.165$ & $0.078$ & $0.157$ & $0.313$ & $0.545$ & $0.790$ \\
        & $(0.481)$ & $(0.513)$ & $(0.563)$ & $(0.478)$ & $(0.479)$ & $(0.495)$ & $(0.519)$ & $(0.539)$ \\
  $1000$ & $0.056$ & $0.108$ & $0.294$ & $0.107$ & $0.282$ & $0.573$ & $0.850$ & $0.980$ \\
        & $(0.480)$ & $(0.510)$ & $(0.565)$ & $(0.472)$ & $(0.483)$ & $(0.494)$ & $(0.517)$ & $(0.546)$ \\
  $2000$ & $0.056$ & $0.165$ & $0.513$ & $0.164$ & $0.504$ & $0.861$ & $0.990$ & $1.000$ \\
        & $(0.480)$ & $(0.506)$ & $(0.561)$ & $(0.473)$ & $(0.478)$ & $(0.492)$ & $(0.518)$ & $(0.544)$ \\ \hline
                 \multicolumn{8}{c}{b) bivariate $t_3$ distribution}\\
  $500$ & $0.051$ & $0.082$ & $0.183$ & $0.084$ & $0.191$ & $0.401$ & $0.671$ & $0.895$ \\
        & $(0.054)$ & $(0.286)$ & $(0.760)$ & $(0.136)$ & $(0.438)$ & $(0.722)$ & $(0.857)$ & $(0.915)$ \\
  $1000$ & $0.053$ & $0.119$ & $0.327$ & $0.122$ & $0.347$ & $0.693$ & $0.933$ & $0.996$ \\
        & $(0.044)$ & $(0.385)$ & $(0.868)$ & $(0.213)$ & $(0.630)$ & $(0.848)$ & $(0.923)$ & $(0.954)$ \\
  $2000$ & $0.055$ & $0.188$ & $0.573$ & $0.193$ & $0.610$ & $0.941$ & $0.999$ & $1.000$ \\
        & $(0.037)$ & $(0.520)$ & $(0.924)$ & $(0.337)$ & $(0.777)$ & $(0.915)$ & $(0.955)$ & $(0.970)$ \\ \hline
               \multicolumn{8}{c}{c) bivariate $t_5$ distribution}\\
  $500$ & $0.053$ & $0.083$ & $0.182$ & $0.087$ & $0.200$ & $0.423$ & $0.696$ & $0.911$ \\
        & $(0.042)$ & $(0.465)$ & $(0.968)$ & $(0.273)$ & $(0.792)$ & $(0.967)$ & $(0.991)$ & $(0.995)$ \\
  $1000$ & $0.054$ & $0.118$ & $0.331$ & $0.124$ & $0.366$ & $0.719$ & $0.947$ & $0.998$ \\
        & $(0.041)$ & $(0.725)$ & $(0.994)$ & $(0.529)$ & $(0.961)$ & $(0.994)$ & $(0.998)$ & $(0.999)$ \\
  $2000$ & $0.055$ & $0.192$ & $0.581$ & $0.199$ & $0.640$ & $0.953$ & $0.999$ & $1.000$ \\
        & $(0.040)$ & $(0.929)$ & $(0.998)$ & $(0.822)$ & $(0.993)$ & $(0.998)$ & $(0.999)$ & $(1.000)$ \\ \hline
        \end{tabular}
\end{table}

\begin{table}[!h!]
\centering
\caption{Empirical power in different settings for the trivariate $t_3$-distribution with serial independence, $\rho_0=0.4$ (for $\rho_1 = -0.6$, the shape matrix would not be positively definite) \label{simupowerthreedim}}
\vspace{0.5cm}
\begin{tabular}{cccccccc}\hline \hline
 $n$ & \multicolumn{7}{c}{Values of $\rho_1$} \\
  & $0.4$ & $0.6$ & $0.8$ & $0.2$ & $0$ & $-0.2$ & $-0.4$ \\ \hline
$500$ & $0.042$ & $0.169$ & $0.534$ & $0.195$ & $0.695$ & $0.988$ & $1.000$ \\
$1000$ & $0.043$ & $0.311$ & $0.850$ & $0.368$ & $0.955$ & $1.000$ & $1.000$ \\
$2000$ & $0.046$ & $0.570$ & $0.992$ & $0.660$ & $1.000$ & $1.000$ & $1.000$ \\ \hline
\end{tabular}
\end{table}

\begin{table}[]
\centering
\caption{Dow Jones and Nasdaq Returns around the Black Monday and empirical moments of the whole time span \label{tablereturns}}
\vspace{0.5cm}
\begin{tabular}{cccccccc}\hline \hline
 Day & 16.10. & 19.10. & 20.10. & 21.10. & 22.10. & $\hat \mu$ & $\hat \sigma$ \\  \hline
Dow Jones & $-0.047$ & $-0.256$ & $0.057$ & $0.097$ & $-0.039$ &
$0.001$ & $0.013$ \\
Nasdaq & $-0.039$ & $-0.120$ & $-0.094$ & $0.071$ & $-0.046$ & $0.001$ & $0.009$\\
\hline
\end{tabular}
\end{table}

\end{document}